\documentclass[11pt]{article}

\usepackage[textwidth=15.2cm,textheight=22cm]{geometry}
\usepackage{amsmath,amssymb}
\usepackage{latexsym}
\usepackage{graphicx}
\usepackage{bbm}
\usepackage{bm}
\usepackage{cite}

\allowdisplaybreaks[1]

\newcommand{\er}{{\rm e}}
\newcommand{\dr}{{\rm d}}
\newcommand{\Tr}{{\rm Tr}}

\newcommand{\R}{\mathbbm{R}}

\newcommand{\be}{\begin{equation}}
\newcommand{\ee}{\end{equation}}
\newcommand{\ba}{\begin{eqnarray}}
\newcommand{\ea}{\end{eqnarray}}
\newcommand{\bdm}{\begin{displaymath}}
\newcommand{\edm}{\end{displaymath}}
\newcommand{\ra}{\rangle}
\newcommand{\la}{\langle}

\newcommand\fr[1]{\frac{1}{#1}}

\newcommand{\Dsm}{\,{\raisebox{0.5pt}{$/$} \hspace{-7.5pt} D}}

\def\b{\beta}
\def\a{\alpha}
\def\g{\gamma}

\def\veps{\varepsilon}

\def\d{\delta}

\def\v{\varphi}
\newcommand{\half}{\frac{1}{2}}
\newcommand{\nn}{\nonumber}

\newcommand{\ie}{{\it i.e.\ }}
\newcommand{\eg}{{\it e.g.\ }}
\newcommand{\hepth}[1]{{\tt hep-th/#1}}

\newcommand{\calN}{{\mathcal N}}

\newcommand{\calS}{{\mathcal S}}

\DeclareMathAlphabet{\mathpzc}{OT1}{pzc}{m}{it}
\newcommand\hsp[1]{\hspace*{#1 cm}}
\newcommand\vsp[1]{\vspace*{#1 cm}}
\newcommand{\ndt}{\noindent}
\newcommand{\mb}[1]{\mathbf{#1}}

\newcommand{\nbar}[1]{\overline{#1}}

\def\bea{\begin{eqnarray}}
\def\eea{\end{eqnarray}}
\def\beas{\begin{eqnarray*}}
\def\eeas{\end{eqnarray*}}
\def\sla{\raise.15ex\hbox{$/$}\kern-.57em}

\def\parm{{\partial}_{-}}

\newcommand{\rd}{{\rm d}}

\newcommand{\vcd}{\varphi}

\begin{document}

\begin{titlepage}
\begin{flushright}  
{\small AEI-2007-006 \\
TCDMATH 07--01}
\end{flushright}
\vskip 1cm

\centerline{\Large{\bf {Proof of ultra-violet finiteness for a planar}}} 
\vskip .3cm
\centerline{\Large{\bf {non-supersymmetric Yang--Mills theory}}}

\vskip 1.5cm

\centerline{Sudarshan Ananth$^\dagger$, Stefano Kovacs$^*$ 
and Hidehiko Shimada$^\dagger$}

\vskip .5cm

\centerline{{\it {$^\dagger$ Max-Planck-Institut f\"{u}r 
Gravitationsphysik}}}
\centerline{{\it {$\;$ Albert-Einstein-Institut, Potsdam, Germany}}}

\vskip 0.5cm

\centerline{\em $^*$ School of Mathematics}
\centerline{\em $\;$ Trinity College, Dublin, Ireland}

\vskip 1.5cm

\centerline{\bf {Abstract}}
\vskip .5cm

\noindent 
This paper focuses on a three-parameter deformation of $\calN=4$
Yang--Mills that breaks all the supersymmetry in the theory. We show
that the resulting non-supersymmetric gauge theory is scale invariant,
in the planar approximation, by proving that its Green functions are
ultra-violet finite to all orders in light-cone perturbation theory. 

\vfill

\end{titlepage}

\section{Introduction}

Scale invariant quantum field theories can be viewed as fixed points
of a renormalization group flow in theory space~\cite{KW}. A
particularly interesting class of such theories are those in which the
scale invariance is preserved while continuous parameters are
varied. In such cases the fixed points constitute a manifold. A well
known example in four dimensions is the maximally ($\mathcal{N}=4$)
supersymmetric Yang--Mills (SYM) theory~\cite{N4}. This theory is
conformally invariant~\cite{finN4,SM,BLN2} for any value of the gauge
coupling, so it corresponds to a one-dimensional manifold of fixed
points~\footnote{Strictly speaking, the references~\cite
{finN4,SM,BLN2} prove that $\calN=4$ SYM is scale invariant, but the
theory is believed to be conformally invariant as well.}.

A class of deformations of $\calN=4$ Yang--Mills, referred to as
$\beta$-deformations~\cite{LS}, are expected to preserve the
conformal symmetry while extending the manifold of fixed points to
higher dimensional surfaces. These $\b$-deformed Yang--Mills
theories~\cite{LS,Beta} are characterized, in $\calN=1$ superspace, by
superpotentials of the form
\be
{\rm W}=\int \dr^4x \left[ \int\dr^2\theta \: g\,h
\,\Tr\left(\er^{i\pi\b} \Phi^1\Phi^2\Phi^3 - \er^{-i\pi\b}
\Phi^1\Phi^3\Phi^2\right) + {\rm h.c.} \right] ,
\label{genbdef}
\ee
where $g$ is the Yang--Mills coupling and $h$ and $\b$ are two
complex parameters. In~\cite{AKS1}, it was shown that the model with
$h=1$ and $\b\in\R$ is scale invariant to all orders in planar
perturbation theory.

The recent proposal of a supergravity solution~\cite{LM} dual to these
deformed models serves as additional motivation to study these
theories in the context of the AdS/CFT correspondence~\cite{adscft}.

In this paper we will focus on the case of a non-supersymmetric
Yang--Mills theory. We will prove that this theory, obtained by
deforming $\calN=4$ Yang--Mills, is scale invariant in the planar
approximation. We will achieve this by proving that planar Green
functions in the theory are ultra-violet finite to all orders in
light-cone perturbation theory. This in turn implies that the
deformed non-supersymmetric gauge theory is scale invariant in the
planar limit. Scale invariance, unlike finiteness of Green functions,
is a gauge-independent statement~\cite{ms}.

The non-supersymmetric (deformed) Yang--Mills theory~\cite{SF,FRT}
will be obtained from $\calN=4$ SYM by a simple generalization of the
superspace $\star\,$-product introduced in~\cite{AKS1}. Realizing the
deformation in this manner will allow us to prove the ultra-violet
finiteness of the theory using the same arguments presented
in~\cite{AKS1}.

Although the deformation breaks supersymmetry completely, there is a
close relation between the perturbative expansions in the deformed
theory and in $\calN=4$ SYM. Specifically, as in the case of the
$\calN=1$ supersymmetric deformation of~\cite{LM}, single planar
diagrams in the deformed theory differ from the corresponding
diagrams in $\calN=4$ SYM only by phase factors. However, it is not
clear whether the finiteness of the deformed model can be deduced
from this fact alone. Individual diagrams in $\calN=4$ SYM can be
divergent and the finiteness of the complete Green functions is, in
general, the result of cancelations among divergent diagrams. In the
deformed theory these cancelations may be ruined if the phase factors
acquired by different diagrams are not the same. This is particularly
relevant in the non-supersymmetric case under consideration since
this theory does not have a formulation which makes available the
powerful techniques of $\calN=1$ superspace. 

Our approach, instead, involves directly proving the ultra-violet
finiteness of the deformed theory without relying on the finiteness
of the original model. We will explicitly show that all Green
functions in the deformed theory are indeed finite.

\section{The non-supersymmetric Yang--Mills theory}

The $\calN=4$ Yang--Mills action in manifestly SU(4) notation is
\ba
\calS&=&\int \dr^4x\,\Tr {\biggl (} \frac{1}{2} F^{\mu\nu}F_{\mu\nu} 
+ D^\mu\bar\v_{mn} D_\mu\v^{mn} - 2i\,\lambda^m \Dsm \bar
\lambda_m \nn \\
&& -2\sqrt{2}\,g \left([\lambda^m,\lambda^n] \bar\v_{mn} 
+[\bar\lambda_p,\bar\lambda_q] \v^{pq}\right)-
\frac{1}{2} g^2 [\v^{mn},\v^{kl}][\bar\v_{mn},\bar\v_{kl}] {\biggr )}\, ,
\label{Ldef} 
\ea
where $m,n=1,\ldots,4$ are indices in the fundamental of SU(4). The
field content consists of a gauge field, $A_\mu$, six real scalars,
introduced as SU(4) bispinors, $\v^{mn}$, satisfying
\be 
\v^{mn}=-\v^{nm}\, ,\qquad \bar\v_{mn}=(\v^{mn})^\dagger=
\frac{1}{2}\,\epsilon_{mnpq}\,{\vcd^{pq}}\, , 
\label{realcond}
\ee
and four Weyl fermions, $\lambda^m$, and their conjugates. All the
fields lie in the adjoint representation of the gauge group. 

The non-supersymmetric deformation involves three real parameters,
$\g_i$, $i=1,2,3$. The deformed theory has the same field content as
$\calN=4$ SYM and is obtained by modifying the Yukawa and scalar
quartic couplings in the $\calN=4$ SYM action by certain phase
factors. 

These phases completely break the supersymmetry as well as the SU(4)
R-symmetry of $\calN=4$ SYM. The resulting theory is invariant under
the Cartan subgroup of SU(4), U(1)$\times$U(1)$\times$U(1), which
arises as a flavor symmetry. The six real scalars and the four Weyl
fermions are charged under this U(1)$\times$U(1)$\times$U(1) symmetry,
while the gauge field remains uncharged.

The phase factors in the action of the deformed theory can be
generated via a $*$-product which generalizes the one introduced
in~\cite{LM} to realize a one-parameter $\calN=1$ $\b$-deformation of
$\calN=4$ SYM. To define the $*$-product we make a choice of basis for
the Cartan subalgebra of SU(4). This corresponds to the assignment of
charges, ${q_1, q_2,q_3}$, given in
table~\ref{compcharges}~\footnote{The charges of the conjugate
fermions, $\bar\lambda_m$, are the opposite of those of the
$\lambda^m$'s. Notice that, working with the bi-spinor representation
for the scalars, the charges of $\v^{mn}$ for any $m$, $n$ pair are
the same as those of the combination $\lambda^m\lambda^n$.}. From the
table it is also clear that the deformed theory cannot preserve any
supersymmetries, for generic deformation parameters, because $A_ \mu$
and $\lambda^4$, which lie in the same $\calN=1$ multiplet, have
different charges.
\begin{table}[!htb]
\begin{center}
\begin{tabular}{|c||c|c|c|}
\hline
\raisebox{-3pt}{\rule{0pt}{14pt}} 
& $q_1$ & $q_2$ & $q_3$ \\
\hline
$\v^{14}$ \raisebox{-3pt}{\rule{0pt}{14pt}} & $1$ & $0$ & $0$ \\
\hline
$\v^{24}$ \raisebox{-3pt}{\rule{0pt}{14pt}} & $0$ & $1$ & $0$ \\
\hline
$\v^{34}$ \raisebox{-3pt}{\rule{0pt}{14pt}} & $0$ & $0$ & $1$ \\
\hline
$\v^{23}$ \raisebox{-3pt}{\rule{0pt}{14pt}} & $-1$ & $0$ & $0$ \\
\hline
$\v^{13}$ \raisebox{-3pt}{\rule{0pt}{14pt}} & $0$ & $-1$ & $0$ \\
\hline
$\v^{12}$ \raisebox{-3pt}{\rule{0pt}{14pt}} & $0$ & $0$ & $-1$ \\ \hline
\end{tabular}
\hsp{0.7}
\begin{tabular}{|c||c|c|c|}
\hline
\raisebox{-3pt}{\rule{0pt}{14pt}} 
& $q_1$ & $q_2$ & $q_3$ \\
\hline
$\lambda^1$ \raisebox{-5pt}{\rule{0pt}{16pt}} & $\half$ & $-\half$ 
& $-\half$ \\ \hline
$\lambda^2$ \raisebox{-5pt}{\rule{0pt}{16pt}} & $-\half$ & $\half$ 
& $-\half$ \\ \hline
$\lambda^3$ \raisebox{-5pt}{\rule{0pt}{16pt}} & $-\half$ & $-\half$ 
& $\half$ \\ \hline
$\lambda^4$ \raisebox{-5pt}{\rule{0pt}{16pt}} & $\half$ & $\half$ 
& $\half$ \\ \hline
$A_\mu$ \raisebox{-5pt}{\rule{0pt}{16pt}} & 0 & 0
& 0 \\ \hline
\end{tabular}
\end{center}
\caption{Flavor charges of the fields in the three-parameter
deformation of $\calN=4$ SYM.}
\label{compcharges}
\end{table}

\ndt
In terms of these charges the $*$-product that realizes the
deformation is~\cite{SF,FRT}
\be
f*g = \er^{i\pi\g_i\veps^{ijk}\,q^f_jq^g_k}fg \, ,
\label{*def}
\ee
where $f$ and $g$ denote two generic component fields. Although the
choice of $q_1$, $q_2$ and $q_3$ is arbitrary, the connection
between the phase factors introduced by the $*$-product (\ref{*def})
and the three charges has a natural interpretation in the dual
supergravity~\cite{LM,SF,FRT,ACO}. The special case in which the
three $\g_i$ parameters are equal corresponds to the $\calN=1$
$\b$-deformed theory of~\cite{LM}.

The deformed non-supersymmetric theory is simply obtained by
replacing all commutators in the $\calN=4$ SYM action (\ref{Ldef}) by
$*$-commutators defined as
\be
[\,f\,,\,g\,]_* = f*g-g*f \, .
\label{*comm} 
\ee 
The component action describing the deformed theory is therefore
\ba
\calS&=&\int \dr^4x\,\Tr {\biggl (} \frac{1}{2} F^{\mu\nu}F_{\mu\nu} 
+ D^\mu{\bar\v}_{mn} D_\mu\v^{mn} - 2i\,\lambda^m \Dsm \bar
\lambda_m \nn \\
&&-2\sqrt{2}\,g \left([\lambda^m,\lambda^n]_*\bar\v_{mn} +
[\bar\lambda_p,\bar\lambda_q]_*\v^{pq}\right)-\frac{1}{2} 
g^2 [\v^{mn},\v^{kl}]_*[\bar\v_{mn},\bar\v_{kl}]_* {\biggr )}\ .
\label{*Ldef} 
\ea

There is an another way to introduce the same $*$-product. We will
find this alternate definition useful since it closely resembles that
used in~\cite{AKS1}. This will allow us to carry over many of the
techniques used there as well.

We choose to define the $*$-product by 
\be
f*g=\er^{i\pi\b(Q^{[1]}_fQ^{[2]}_g-Q^{[2]}_fQ^{[1]}_g)}\,f\,g \, ,
\label{new*prod}
\ee 
where $\b$ is a real parameter and $Q^{[1]}$ and $Q^{[2]}$ are two
commuting generators of SU(4) which can be represented by diagonal
matrices
\begin{equation}
\left( \begin{array}{cccc}
\a^{[1]}_1 & & & \\
&\a^{[1]}_2 & & \\
& &\a^{[1]}_3 & \\
& & &\a^{[1]}_4 \\
\end{array} \right) \, , \qquad
\left( \begin{array}{cccc}
\a^{[2]}_1 & & & \\
&\a^{[2]}_2 & & \\
& &\a^{[2]}_3 & \\
& & &\a^{[2]}_4 \\
\end{array} \right) \, .
\end{equation}
These matrices act on the $\mb4$ of SU(4) so the $Q^{[1]}$ charge of
$\lambda^m$, for example, is $\a^{[1]}_m$ while the $Q^{[2]}$ charge
of $\v^{mn}$ is $\a^{[2]}_m+\a^{[2]}_n$ (see footnote~$2$). The $\a$'s
parametrizing the two U(1) generators are constrained by the
tracelessness conditions
\be
\label{constr}
\sum_{m=1}^4 \alpha^{[1]}_m = 0 \, , \qquad 
\sum_{m=1}^4 \alpha^{[2]}_m = 0 \, .
\ee
The effect of the $*$-product is best illustrated by computing
explicitly the phases that it introduces in the products of
fields. For example, in the commutator of the Weyl fermions one
obtains
\be
[\lambda^m,\lambda^n]_*= \er^{i\pi\b_{mn}}\lambda^m \lambda^n - 
\er^{i\pi\b_{nm}}\lambda^n \lambda^m \, ,
\ee
where
\be
\b_{mn}=\b (\alpha^{[1]}_m \alpha^{[2]}_n - 
\alpha^{[2]}_n \alpha^{[1]}_m) \, , \qquad \b_{nm}=-\b_{mn} \, .
\ee
The $*$-commutators of the scalars (expressed as SU(4) bi-spinors)
can be similarly computed. Note that all the $*$-products involving
the gauge field, which is a SU(4) singlet, reduce to ordinary
products. 

Equation (\ref{constr}) implies that the $\b_{mn}$ parameters are not
all independent. For example \be \b_{14}=-\b_{24}-\b_{34} \, . \ee
It is easy to verify that the phases introduced by the $*$-product in
the deformed action can all be written in terms of just three
parameters, \eg $\b_{12}$, $\b_{23}$ and $\b_{31}$, using the
tracelessness constraints (\ref{constr}). Thus the deformed model
represents a three-parameter non-supersymmetric deformation of
$\calN=4$ Yang--Mills. With this definition of the $*$-product the
special case of the $\calN$=1 supersymmetric deformation of~\cite{LM}
corresponds to the choice $\a^{[1]}_4=\a^{[2]}_4=0$.

It is straightforward to verify the equivalence of the two definitions
of the $*$-product in (\ref{*def}) and in (\ref{new*prod}). To find
the relation between the parameters used in the two cases we expand
the two charges, $Q^{[1]}$ and $Q^{[2]}$, in the basis formed by
$q_1, q_2$ and $q_3$, 
\begin{equation}
Q^{[r]}= (\alpha^{[r]}_1+\alpha^{[r]}_4)\, q_1 
+(\alpha^{[r]}_2+\alpha^{[r]}_4)\, q_2 
+(\alpha^{[r]}_3+\alpha^{[r]}_4)\, q_3 \, ,
\qquad r=1,2 \, .
\label{Qqrel}
\end{equation}
By substituting this expression into (\ref{new*prod}), we see that it
coincides with the definition (\ref{*def}), if we set
\begin{eqnarray}
\gamma_1&\!\!=\!\!&\beta \left((\alpha^{[1]}_2+\alpha^{[1]}_4)(\alpha^
{[2]}_3
+\alpha^{[2]}_4)-(\alpha^{[1]}_3+\alpha^{[1]}_4)(\alpha^{[2]}_2
+\alpha^{[2]}_4) \right) \nn \\
\gamma_2&\!\!=\!\!&\beta \left((\alpha^{[1]}_3+\alpha^{[1]}_4)(\alpha^
{[2]}_1
+\alpha^{[2]}_4)-(\alpha^{[1]}_1+\alpha^{[1]}_4)(\alpha^{[2]}_3
+\alpha^{[2]}_4) \right) \label{alphagamma}\\
\gamma_3&\!\!=\!\!&\beta \left((\alpha^{[1]}_1+\alpha^{[1]}_4)(\alpha^
{[2]}_2
+\alpha^{[2]}_4)-(\alpha^{[1]}_2+\alpha^{[1]}_4)(\alpha^{[2]}_1
+\alpha^{[2]}_4) \right) \, . \nn
\end{eqnarray}
Having established the equivalence of the two $*$-products, in the
following we will work with the definition in (\ref{new*prod}).

\section{$\calN=4$ light-cone superspace}

In the proof of finiteness we will use the tools of light-cone
superspace~\cite{BLN2,SM}. Despite the fact that the deformed
Yang--Mills theory is non-supersymmetric it can still be formulated in
$\calN=4$ light-cone superspace. This is thanks to the fact that its
field content is identical to that of $\calN=4$ Yang--Mills. Thus as a
first step towards a light-cone superspace realization we formulate
the deformed non-supersymmetric theory in (\ref{*Ldef}) in light-cone
gauge.

The choice of light-cone gauge is made by setting
\beas
A_-=0\ .
\eeas
The $A_+$ component is solved for using the equations of motion. The
SU(4) fermions split up as
\beas
\lambda^m_\a\rightarrow\,(\lambda^{m(+)},\lambda^{m(-)})\ .
\eeas
Again, the equations of motion allow us to eliminate
$\lambda^{m(+)}$. For simplicity of notation we rename the remaining
physical field to $\chi^m$. 

We then derive the light-cone component description of the deformed
non-supersymmetric theory applying these steps to the action (\ref
{*Ldef}). The light-cone non-supersymmetric theory can also be
obtained by replacing all the commutators of charged fields (all six
scalars and four fermions) in the $\calN=4$ light-cone component
action~\cite{BLN1} by $*$-commutators. The exact form of the
component action is irrelevant to this paper as, in the following, we
will use the light-cone superspace formalism. We refer the reader
to~\cite{AKS1} for further details regarding the light-cone component
description.

The $\calN=4$ light-cone superspace~\cite{BLN1,BBB2,SA1} is
comprised of four bosonic coordinates, $x^+,x^-,x,\bar x$, and eight
fermionic coordinates, $\theta^m,{\bar \theta}_m$, $m=1,2,3,4$. These
are collectively denoted by $z=(x^+,x^-,x,{\bar x},\theta^m,{\bar
\theta}_m)$. 

All the degrees of freedom of the deformed theory are described by a
single scalar superfield. This superfield is defined by the chirality
constraints
\be
\label{chir}
d^m\,\Phi=0\,,\qquad {\bar d}_n\,{\bar \Phi}=0\,, 
\ee
as well as the ``inside-out constraints''
\be
\label{io}
{\bar d}_m{\bar d}_n\Phi=\frac{1}{2}\epsilon_{mnpq}d^pd^q
{\bar \Phi}\, ,
\ee
where $\bar \Phi$ is the complex conjugate of $\Phi$. The superspace
chiral derivatives in the above expressions are
\be
\label{scd}
{d^{\,m}}=-\frac{\partial}{\partial {\bar \theta}_m}\,+\,
\frac{i}{\sqrt 2}\,{\theta^m}\,
{\partial_-}\,,\qquad{{\bar d}_{\,n}}=\frac{\partial}{\partial \theta^n}\,
-\,\frac{i}{\sqrt 2}\,{{\bar \theta}_n}\,{\partial_-}\ .
\ee
The superfield satisfying the constraints (\ref {chir}) and (\ref{io})
is~\cite{BLN1}
\begin{align}
\label{superf}
\Phi(x,\theta,\bar\theta)&=-\frac{1}{\parm}A(y)-\frac{i}{\parm}\theta^m
{\bar \chi}_m(y)+\frac{i}{\sqrt 2}\theta^m\theta^n
{\nbar \vcd}_{mn}(y) \nn \\
&+\frac{\sqrt 2}{6}\theta^m\theta^n\theta^p\epsilon_{mnpq}
\chi^q(y)-\frac{1}{12}\theta^m\theta^n\theta^p\theta^q
\epsilon_{mnpq}\parm\,{\bar A}(y)\ , 
\end{align}
where $y=\,(\,x,\,{\bar x},\,{x^+},\,y^-\,\equiv\,{x^-}-\,
\frac{i}{\sqrt 2}\,{\theta^m}\,{\bar \theta}_m\,)$ is the chiral
coordinate and the r. h. s. of (\ref{superf}) is to be understood as
an expansion around $x^-$. Note that we use, for the $\frac{1}{\parm}
$ operator, the prescription given in~\cite{SM}.

We now introduce a superspace $\star$-product whose effect on
superfields mimics the action of the $*$-product on component
fields~\cite{AKS1}. In superspace we formally consider the U(1)
generators to be acting on the superspace fermionic coordinates, \ie
we think of the ``flavor'' charges as being carried by the $\theta$
variables (the charges of the ${\bar \theta}$'s are opposite to those
of the $\theta$'s). The superspace $\star$-product is then simply
realized in terms of operators which count the number of $\theta$'s
and $\bar\theta$'s. 

In superspace, we define charges, $q_m$, by
\bea
\label{charges}
&&\hsp{-2}{\overrightarrow q}_{\!m}=\theta^m\frac{{\overrightarrow
{\partial}}}{\partial\theta^m}-\bar\theta_m\frac{{\overrightarrow
{\partial}}}{\partial\bar\theta_m}\ , \nn \\
&&\hsp{-2}{\overleftarrow q}_{\!m}={\frac{{\overleftarrow
{\partial}}}{\partial\theta^m}}\theta^m-\frac{{\overleftarrow
{\partial}}}{\partial\bar\theta_m}\bar\theta_m\ .
\eea
In terms of these we define $Q^{[1]}$ and $Q^{[2]}$ by
\be
\label{char}
Q^{[1]} = \sum_{m=1}^4 \alpha^{[1]}_m \, q_m\ , \qquad
Q^{[2]} = \sum_{m=1}^4 \alpha^{[2]}_m \, q_m\ , 
\ee
where the parameters $\a^{[1]}_m$ and $\a^{[2]}_m$ are the same as
those used for the component $*$-product. We define the
$\star$-product of two superfields, $F$ and $G$, by
\be
\label{sstar}
F \star G = F\,\er^{i\pi{\b}({\overleftarrow Q}^{[1]}_F
{\overrightarrow Q}^{[2]}_G-{\overleftarrow Q}^{[2]}_F
{\overrightarrow Q}^{[1]}_G)}\,G\ 
\ee
and the associated $\star$-commutator by
\be
[\,F\,,\,G\,]_\star = F\star G-G\star F \, .
\label{starcomm}
\ee
The $\star$-product (\ref{sstar}) allows us to formulate the non-
supersymmetric deformed Yang--Mills theory in $\calN=4$ light-cone
superspace. The action reads~\footnote{We remind the reader that as
far as space-time is concerned this is an ordinary ``commutative''
field theory.}
\bea
\label{ans}
S&\!\!\!=\!\!\!& 72\int\dr^4x \int\dr^4\theta\,
{\rd}^4{\bar \theta}\;\Tr\left\{-2\,{\bar \Phi}\,
\frac{\Box}{\parm^2}\,\Phi
+i\,\frac{8}{3}\,g \left(\frac{1}{\parm}{\bar \Phi}\,
[\Phi,{\bar \partial}\Phi]_\star\,+\,\frac{1}{\parm}\,\Phi\,
[{\bar \Phi},\partial{\bar \Phi}]_\star\right) \right. \nn \\
&& \hsp{1}\left. +\,2\,g^2\left(\frac{1}{\parm}[\Phi,\parm\Phi]_\star\,
\frac{1}{\parm}[{\bar \Phi},\parm{\bar \Phi}]_\star\,
+\half[\Phi,{\bar \Phi}]_\star\,[\Phi,{\bar \Phi}]_\star 
\right)\right\} \, .
\eea
Expanding the various $\star$-commutators and performing the
Grassmann integrations reproduces exactly the light-cone component
action justifying our definition of the superspace $\star$-product.

We denote the superfield propagator by 
\bea
\la\Phi^u_{\;\;v}(z_1)\,\Phi^r_{\;\;s}(z_2)\ra = 
\Delta^{u\;r}_{\;\;v\;s}\,(z_1-z_2)\, ,
\eea
where we have made explicit the matrix indices. The corresponding
momentum-space propagator is
\bea
\label{prop}
\Delta^{u\,r}_{\;\;v\,s}(k,\theta_{(1)},{\bar \theta}_{(1)},\theta_{(2)},
{\bar \theta}_{(2)})\,=\,t^{u\,r}_{\;\;v\,s}\,\fr{k_\mu^2}\,d_{(1)}^4\,
\delta^8(\theta_{(1)}-\theta_{(2)})\, ,
\eea
where $\theta_{(1)}$ and $\theta_{(2)}$ denote the fermionic
coordinates at superspace points $z_1$ and $z_2$ respectively and
$t^{u\,r}_{\;\;v\,s}$ is a tensor whose precise structure depends on
the choice of gauge group and is irrelevant to our analysis. Notice
that here and in the following we denote the product of four chiral
derivatives, $d^1d^2d^3d^4$, by $d^4$ and the product of four
anti-chiral derivatives, ${\bar d}_1{\bar d}_2{\bar d}_3{\bar d}_4$,
by ${\bar d}^4$.

The fermionic $\delta$-function is
\be
\delta^8(\theta_{(1)}-\theta_{(2)})=\left(\theta_{(1)}-\theta_{(2)}
\right)^{\!4} \left(\bar\theta_{(1)}-\bar\theta_{(2)}\right)^{\!4} \, .
\label{fdelta}
\ee
The light-cone superspace Feynman rules for the theory can be easily
derived from the action~(\ref{ans}). Having formulated the deformed
theory in $\calN=4$ light-cone superspace we will now prove its
ultra-violet finiteness to all orders in planar perturbation theory.

\section{Proof of finiteness}

The proof of finiteness is identical to that presented
in~\cite{AKS1}. We therefore simply highlight the salient points
involved and refer the reader to~\cite{AKS1} for a complete and
detailed description of the procedure.

At the basis of the proof is Weinberg's power counting theorem~\cite
{SW}, which states that an arbitrary Feynman diagram is convergent if
the superficial degree of divergence, $D$, of the diagram as a whole
and of all its sub-diagrams is negative. To prove the finiteness of
the non-supersymmetric deformed theory we will show that, in
light-cone superspace, all the supergraphs in the theory satisfy the
hypotheses of Weinberg's theorem. This is achieved in two steps: 
\begin{itemize}
\item[$i)$] A superspace dimensional analysis provides a preliminary
estimate which yields $D=0$ for a generic supergraph;
\item [$ii)$] Using superspace manipulations it is shown that the
degree of divergence of any sub-graph in an arbitrary supergraph can
be reduced to a negative value by bringing factors of momentum out of
the internal loops.
\end{itemize}

The first step is based on a version of the superspace power counting
methods of~\cite{GRS} adapted to the light-cone formalism. An
essential ingredient in this analysis is the relation
\be
\d^8(\theta_{(1)}-\theta_{(2)})\,d_{(1)}^4\,\bar d_{(1)}^4\,
\d^8(\theta_{(1)}-\theta_{(2)})=\d^8(\theta_{(1)}-\theta_{(2)})\, .
\label{grsrel}
\ee
A modified version of this equation~\cite{AKS1} remains valid for
planar diagrams in the deformed theory. Repeating the analysis
of~\cite{AKS1} leads to the conclusion that $D=0$ for any planar
supergraph if all the momenta are assumed to contribute to the loop
integrals. 

This estimate is then refined by distinguishing between internal and
external lines and using the explicit form of the vertices
in~(\ref{ans}): standard manipulations in light-cone superspace allow
us to show that the degree of divergence of any loop in an arbitrary
supergraph is actually negative. 

\begin{figure}[!hbt]
\begin{center}
\includegraphics[width=0.35\textwidth]{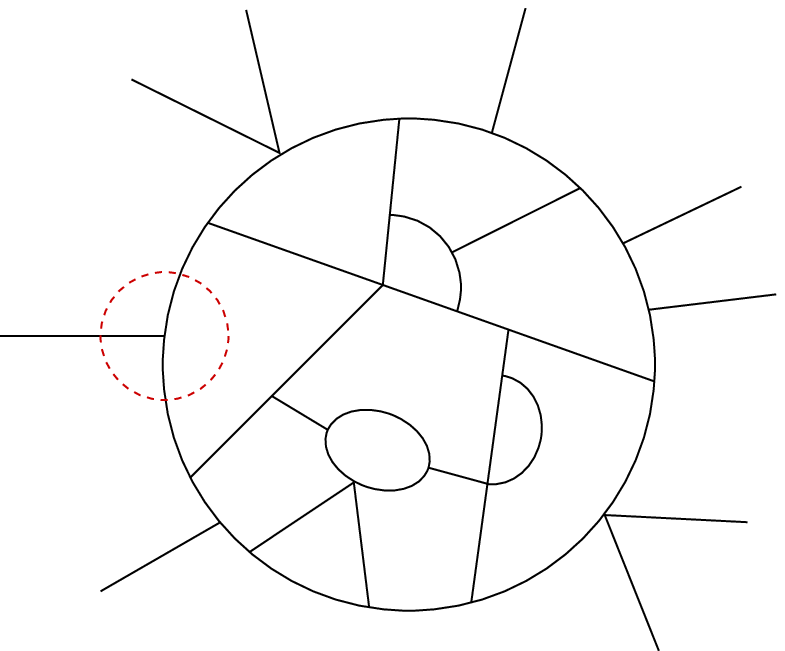}
\hsp{2}
\includegraphics[width=0.35\textwidth]{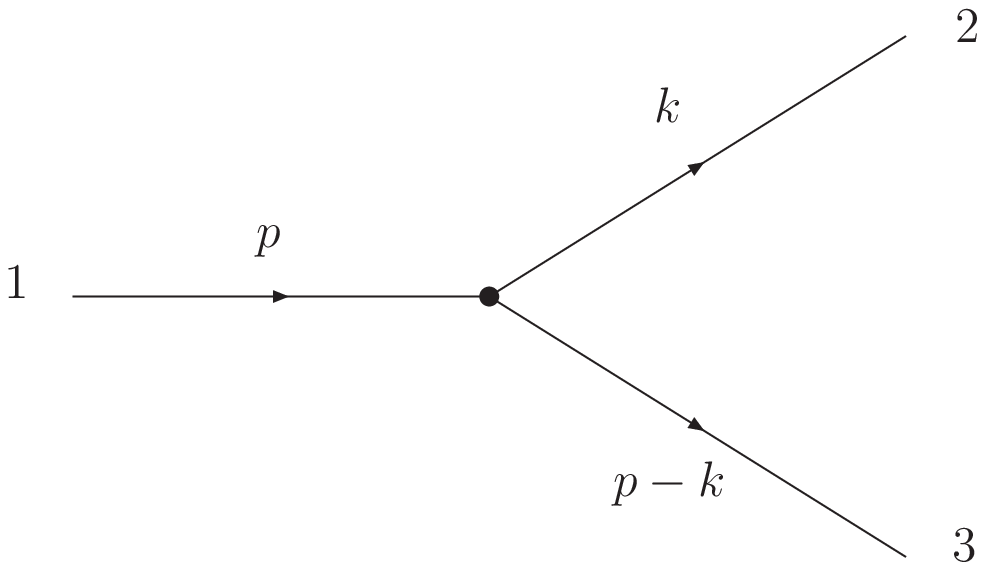}
\end{center}
\vsp{-0.3}
\caption{Light-cone supergraphs}
\label{fig}
\end{figure}

The strategy used in the analysis of a generic complicated
supergraph, such as the one depicted on the left hand side of
figure~\ref{fig}, is the following. 

We consider an external leg, \eg the one marked with a dotted circle
in the figure, and analyze the internal loop it connects to. The
corresponding integral can be rendered finite using integrations by
parts in superspace to move chiral derivatives from the internal lines
onto the external one. This brings factors of momentum out of the loop
integral and improves its convergence~\footnote{In general to cancel
the divergent part of a graph it is necessary to combine the
contribution of different Wick contractions.}. External lines
connected to a quartic vertex can be analyzed in a similar
fashion. Once the first loop connected to the selected external leg
is rendered finite, we move to an adjacent loop and use similar
manipulations to reduce its degree of divergence to a negative
value. The procedure continues until all the loops in the supergraph
have been dealt with. 

In order to give the reader a flavor of the kind of manipulations
involved, we analyze explicitly the contribution of specific Wick
contractions to the loop attached to the vertex highlighted in the
graph on the left of figure \ref{fig}. We isolate the vertex, which is
depicted on the right hand side of the figure with the associated
momenta. In position space we consider
\bea
\la\Phi(z_1)\Phi(z_2)\Phi(z_3) \int\!\dr^{12}z\:i{\cal L}_3(z)\ra \, ,
\label{3ptf}
\eea
where
\bea
i\,{\cal L}_3\,(z) = -g \,\Tr\,{\biggl [}\frac{1}{12}\,\fr{\parm}\,
\Phi\,[{\frac{{\bar d}^4}{\parm^2}\Phi},\partial
{\frac{{\bar d}^4}{\parm^2}\Phi}]_\star{\biggr ]} .
\eea
Among the Wick contractions contributing to (\ref{3ptf}) we consider 
\bea
\Tr\left\{\fr{\parm}{\Delta_2}[\frac{{\bar d}^4}{\parm^2}\Delta_1,
\partial\,\frac{{{{\bar d}^4}}}{\parm^2}\Delta_3]_\star+\fr{\parm}\Delta_3
[\frac{{\bar d}^4}{\parm^2}\Delta_1,\partial\,
\frac{{{\bar d}^4}}{\parm^2}{\Delta_2}]_\star\right\}\, ,
\label{wcontr}
\eea
where the index 1 refers to the external leg while 2 and 3 identify
the internal legs. In the above formula the propagators
$\Delta_n\equiv \Delta(z-z_n)$ are treated as matrices with indices
associated with the interaction point $z$ (indices relating to $z_n$,
$n=1,2,3$, are omitted).

The manipulations required to render a supergraph finite are exactly
the same as those used in the proof of finiteness of $\calN$=4 SYM
in~\cite{SM, BLN2}. An identical analysis was applied to the case of
the $\calN$=1 $\b$-deformed theory in~\cite{AKS1}, where it was
shown that the arguments in the $\calN$=4 proof remain applicable
thanks to the properties of the superspace
$\star$-product~\footnote{For a detailed list of properties of the
$\star$-product we refer the reader to appendix A
of~\cite{AKS1}.}. The same can be shown in the present
non-supersymmetric case. In general in the presence of
$\star$-products the integration by parts of chiral derivatives
introduces phase factors in the superspace expressions, however, as
discussed in~\cite{AKS1}, these do not affect the proof. 

In order to prove that the contribution of the contractions in (\ref
{wcontr}) to the loop integral is ultra-violet finite, we integrate
the superspace chiral derivatives from leg $3$ to leg $2$ in the
first term. Using the associativity of the $\star$-product and the
cyclicity of the trace, the result can be rewritten as 
\bea
\Tr\left\{-\partial\,\fr{\parm^2}\Delta_3[
\frac{{\bar d}^4}{\parm^2}\Delta_1,
\frac{{\bar d}^4}{\parm}\Delta_2]_\star+\fr{\parm}\Delta_3
[\frac{{\bar d}^4}{\parm^2}\Delta_1,\partial\,
\frac{{\bar d}^4}{\parm^2}\Delta_2]_\star \right\} \, , 
\eea
so that a common structure can be factored out. In momentum space 
we get
\bea
-\,{\biggl (}\frac{p-k}{{(p_--k_-)}^2}\,\fr{{p_-}^2}\,
\fr{k_-}\,-\,\fr{p_--k_-}\,\fr{{p_-}^2}\,\frac{k}{{k_-}^2}\,{\biggr )}\,
\Tr\left(\Delta_3\,[{\bar d}^4\Delta_1,{{\bar d}^4}{\Delta_2}
]_\star\right) \, .
\eea
Potential ultra-violet divergences arise from loop momenta, $k$,
satisfying $k\,\gg\,p$. In this limit the leading terms in parentheses
cancel. This means that the logarithmically divergent contribution
vanishes leaving a finite integral. 

All the other Wick contractions involving both cubic and quartic
vertices can be treated following similar steps. This leads to the
conclusion that all the loops connected to external legs in an
arbitrarily complicated supergraph have negative degree of
divergence. We can then proceed to internal loops and repeat the
same analysis. The ultra-violet finiteness of any supergraph then
follows from the application of Weinberg's theorem~\cite{SW}. Note
that the use of the theorem in the light-cone gauge is permitted due
to our choice of pole structure~\cite{SM}, which allows for Wick
rotation into Euclidean space. 

\vskip 0.65 cm

\begin{center}
* ~ * ~ *
\end{center}

\vskip -0.1 cm

Non-planar diagrams can be analyzed using the same methods. However,
in the non-planar case the relation (\ref{grsrel}) does not hold,
implying that the preliminary estimate yielding $D=0$ for all
supergraphs is no longer valid. Therefore manipulations of the type
outlined above are not sufficient to conclude that generic non-planar
diagrams have negative superficial degree of divergence. 

As already observed, it is not straightforward to deduce the
finiteness of the non-supersymmetric deformation of $\calN=4$ SYM
considered in this paper from the ultra-violet properties of the
parent $\calN=4$ theory. We also point out that, while there exist
other indirect arguments for the finiteness of theories obtained as
deformations of $\calN$=4 SYM, such as those described
in~\cite{LS,oldfinite}, these methods rely on supersymmetry. Hence
they are not applicable to non-supersymmetric theories such as the one
studied here. For these theories the light-cone analysis presented
here is so far the only viable approach. 

There are natural generalizations of the deformation considered in
this paper. Mass terms for all the matter fields can be added
preserving ultra-violet finiteness, but at the expense of scale
invariance. Moreover in the light-cone superspace formulation
additional parameters can be added multiplying respectively the two
cubic couplings and the two quartic couplings in the
action~(\ref{ans}). This does not ruin the proof of scale invariance,
since in proving the finiteness of the light-cone Green functions only
cancelations among supergraphs involving separately the cubic and
quartic vertices were invoked. However, this type of deformation will
probably break Lorentz invariance. A more interesting generalization
involves making the deformation parameters complex. A supergravity
solution with 2+6 parameters, which generalizes the supergravity dual
to the deformed theory considered in this paper, was obtained
in~\cite{SF}. We expect that the techniques used here and
in~\cite{AKS1} will allow us to prove the finiteness, in the planar
approximation, of the theories deformed with complex parameters. This
case is of particular interest in connection with the recent work on
the role of integrability in the context of the gauge/gravity
correspondence. The $\calN=1$ deformation of~\cite{LM} and the
non-supersymmetric case studied in this paper are believed to preserve
the integrability of the spectrum in the planar approximation only in
the case of real deformation parameters~\cite{BC,SF,FRT}.
Understanding whether the complex deformations indeed lead to finite
theories may help to shed light on the interconnections between
integrability and scale invariance. 

Finally, the maximally supersymmetric $\calN=8$ supergravity in four
dimensions~\cite{CJS} has also been formulated in light-cone
superspace up to second order in the gravitational coupling
constant~\cite{SA1,BBB}. The main feature of this formulation, as in
the Yang--Mills case, is that it is free of both auxiliary fields and
ghosts. It will be interesting to investigate whether the techniques
presented in this paper and in~\cite{SM,BLN2,AKS1} prove useful in the
study of the ultra-violet behavior of $\calN=8$
supergravity~\cite{MBG}.

\subsection*{Acknowledgments}
We thank Sergey Frolov and Stefan Theisen for useful discussions. SA
thanks the members of the theory groups at the Harish-Chandra
Research Institute and the Tata Institute of Fundamental Research for
helpful comments. SA also acknowledges the kind hospitality of the
Chennai Mathematical Institute. The work of SK was supported in part
by a Marie Curie Intra-European Fellowship and by the EU-RTN network
{\it Constituents, Fundamental Forces and Symmetries of the Universe}
(MRTN-CT-2004-005104).

\end{document}